\documentclass[11pt]{article}
\usepackage{graphicx}

\usepackage{geometry} 
\begin{document}

\textbf{Asymmetry for tensor }\textbf{\textit{t}}$_{2j}$\textbf{ and vector
}\textbf{\textit{t}}$_{1i}$\textbf{ polarizations with taking into account
the deuteron wave function in coordinate space}

\begin{center}
\textbf{V. I. Zhaba}
\end{center}

\begin{center}
\textit{Uzhgorod National University, Department of Theoretical
Physics,}
\end{center}

\begin{center}
\textit{54, Voloshyna St., Uzhgorod, UA-88000, Ukraine}
\end{center}

\textbf{Abstract}

The analytic forms of deuteron wave function in coordinate space
were applied for theoretical calculations of full set asymmetries
of tensor $t_{2j}$ and vector $t_{1i}$ polarizations.
Nucleon-nucleon realistic phenomenological potentials of Nijmegen
group (Nijm1, Nijm2, Nijm93, Reid93) and Argonne group (Argonne
v18) as well as other widely used and popular potentials (OBEPC,
MT, Paris) are used for numerical calculations. The angular
asymmetry is calculated in the range of momentums 0-7 fm$^{ - 1}$.
The values of the asymmetry $t_{ij,}$ by polarization type and
with each other, are analysed. This is the angular-momentum
dependence of values vector $t_{1i} (p,\theta _e )$ and tensor
$t_{2i} (p,\theta _e )$ polarizations in 3D format at momentums
0-7 fm$^{-1}$ and scattering angles 1-180 degrees were
calculated and compared for Reid93 potential. The perspectives of
further application of the obtained results for calculating the
values of the cross-sections, asymmetries and other
characteristics of processes with the participation of a deuteron
are discussed.



\textbf{1. Introduction}

Deuteron is the simplest core, which consists of the two
elementary particles both a proton and a neutron. The simplicity
and clarity of the deuteron structure always serves as a
convenient laboratory for evident simulation and structural
analysing nucleon-nucleon forces \cite{MPLA2018}.

Deuteron is the possible target for an electron beam or the
particle that dissipates on a proton or atomic nuclei. Some
examples could be presented. The paper \cite{Hasell2011} presents
recent results for spin-dependent scattering of electrons on
polarized protons and deuterons for BLAST experiment in MIT-Bates.
The radiative corrections are investigated for polarization
observed in elastic ed- scattering in leptonic variables by
\cite{Gakh2012}. Global analysis of quasi-elastic ed-scattering
data's in the weak-binding approximation was conducted in
\cite{Ethier2014}.

The experiments were performed on the separate measurement of
deuteron form factors in elastic ed-scattering in the interval of
transmitted momentums $p^{2}$=8-15 fm$^{-2}$ on the electron
positron storage VEPP-3 \cite{Nikolenko2010}. The results of
measuring the components of the analyzing powers $T_{2i}$ in the
photodisintegration reaction of a tensor-polarized deuteron are
presented.

The authors \cite{Nikolenko2010} first measured the analyzing
powers for the coherent photoproduction of a neutral pion $\pi
^{0}$ on a tensor-polarized deuteron. According to
\cite{Darwish2018}, tensor target spin asymmetries are calculated
in coherent $\pi ^{0}$- photoproduction on the deuteron, including
an intermediate $\eta $NN- interaction in the three-particle
approach.

Spin observables in dp-scattering and T-invariance test were
studied in the application of the modified Glauber theory by
\cite{Temerbayev2015}. A complete set of deuteron analyzing powers
in elastic dp-scattering at 190 MeV/nucleon is given in
\cite{Sekiguchi20172}. The proton and deuteron analyzing powers
and 10 spin correlation coefficients were measured for elastic p+d
scattering with the energy for bombarding protons 135 and 200 MeV
\cite{Przewoski2006}. Experimental data's are compared with
Fadeev's calculations for CD-Bonn and AV18 potentials. Vector
$A_{y}$ and tensor $A_{xx}$, $A_{yy}$ analyzing powers in elastic
dp-scattering at deuteron kinetic energy $T_{d}$=1.2 and 2.27 GeV
are obtained using the ANKE spectrometer on the COSY storage ring
\cite{Mchedlishvili2018}. The results are compared with other
experimental data's and predictions within the framework of the
multiple scattering of Glauber theory. The authors of
\cite{Gakh2014} investigated the effects of polarization observed
in elastic lepton-deuteron scattering including the lepton masses.

The angular dependence of the tensor $А_{уу}$ and vector $А_{у}$
analyzing powers in inelastic (d,d')-scattering of deuterons with
a momentum at 9.0 GeV/c on hydrogen and carbon in
\cite{Ladygin2006} and of 5.0 GeV/c on beryllium in
\cite{Azhgirey2005} is measured. Further theoretical calculations
and prospects for the process (d,d') are analysed in
\cite{MPLA2018}.

Generally speaking, in general for the theoretical study of mechanisms,
changes and characteristics for the vast majority of these processes with
the participation of a deuteron it is necessary to know the deuteron wave
function (DWF) in the coordinate or momentum space, as well as the deuteron
form factors.

In the last detailed review \cite{ZhabaArxiv}, the static
parameters of deuteron obtained from DWF for different
nucleon-nucleon potentials and models are systematized, as well as
a review, list and characteristics for analytical forms of DWF in
the coordinate space have been reviewed.

In this paper we use the analytic forms of DWF in coordinate space for
theoretical calculations of set asymmetries of tensor $t_{2j}$ and vector
$t_{1i}$ polarizations. Nucleon-nucleon realistic phenomenological potentials
of Nijmegen group (NijmI, NijmII, Nijm93, Reid93) and Argonne group (Argonne
v18) as well as other widely used and popular potentials (OBEPC, MT, Paris)
are used for numerical calculations.

\textbf{2. DWF in Coordinate Space}

Among the large and varied list of analytical forms for DWF in the
coordinate space, elementary parametrization of DWF for the Paris
potential \cite{Lacombe1981} is worth highlighting

\begin{equation}
\label{eq1}
\left\{ {\begin{array}{l}
 u\left( r \right) = \sum\limits_{j = 1}^N {C_j \exp \left( { - m_j r}
\right),} \\
 w\left( r \right) = \sum\limits_{j = 1}^N {D_j \exp \left( { - m_j r}
\right)\left[ {1 + \frac{3}{m_j r} + \frac{3}{\left( {m_j r} \right)^2}}
\right],} \\
 \end{array}} \right.
\end{equation}

where $m_j = \beta + (j - 1)m_0 $; $\beta = \sqrt {ME_d } $,
$m_{0}$=0.9 fm$^{ - 1}$; $M$ -- nucleon mass; $E_{d}$ -- binding
energy of the deuteron. The asymptotics of the deuteron wave
function (\ref{eq1}) at $r \to 0$:

\[
u\left( r \right) \to r;
\quad
w\left( r \right) \to r^3.
\]

The asymptotics of the deuteron wave function (\ref{eq1}) for values $r \to \infty $

\[
\left\{ {\begin{array}{l}
 u(r) \sim A_S \exp ( - \beta r), \\
 w(r) \sim A_D \exp ( - \beta r)\left[ {1 + \frac{3}{\beta r} +
\frac{3}{(\beta r)^2}} \right], \\
 \end{array}} \right.
\]

where $A_{S}$ and $A_{D}$ are the asymptotics of $S$- and $D$-
state normalizations accordingly.

Usually, coefficients $C_{j}$, $D_{j}$ are indicated in the tables in most
cases for specific potentials, except for the last values of the
coefficients $C_{n}$, $D_{n - 2}$, $D_{n - 1}$, $D_{n }$

The search for the coefficients $C_{j}$, $D_{j}$ of the analytical
form (\ref{eq1}) was done for the Paris \cite{Lacombe1981} and the
Bonn (OBEPC \cite{Machleidt1989} and charge-dependent Bonn
(CD-Bonn) \cite{Machleidt2001}) potentials and the fss2 model
(with the Coulomb exchange kernel \cite{Fujiwara2001}), with
$N$=13, 11 and 11 respectively. The formula (\ref{eq1}) was also
applied to MT model \cite{Krutov2007}, where for radial component
of DWF for both the S-states and D- states of number made
$N_{S}$=16 and $N_{D}$=12 accordingly.

In addition to the DWF (\ref{eq1}), the next analytical form was
suggested in paper \cite{NPAE2016} for Nijmegen group potentials
(NijmI, NijmII and Nijm93) and further was used in paper
\cite{MPLA2016} for approximation of DWFs for the Reid93 and
Argonne v18 potentials:

\begin{equation}
\label{eq2}
\left\{ {\begin{array}{l}
 u(r) = r^{3 / 2}\sum\limits_{i = 1}^N {A_i \exp ( - a_i r^3),} \\
 w(r) = r\sum\limits_{i = 1}^N {B_i \exp ( - b_i r^3).} \\
 \end{array}} \right.
\end{equation}

A complete set of coefficients for DWF (\ref{eq2}) for these five
potentials is given in \cite{NPAE2016, MPLA2016}.

Coefficients of DWF for the dressed dibaryon model (DDM)
\cite{Platonova2010} were obtained for the analytical form
\cite{Krasnopolsky1985}:

\begin{equation}
\label{eq3}
\left\{ {\begin{array}{l}
 u(r) = r\sum\limits_{i = 1}^N {A_i \exp ( - a_i r^2),} \\
 w(r) = r^3\sum\limits_{i = 1}^N {B_i \exp ( - b_i r^2).} \\
 \end{array}} \right.
\end{equation}

The new analytical DWFs were also proposed in coordinate space in
such a simple form \cite{Zhaba20162} in 2016:

\begin{equation}
\label{eq4}
\left\{ {\begin{array}{l}
 u(r) = r\sum\limits_{i = 1}^N {A_i \exp ( - a_i r^2),} \\
 w(r) = r\sum\limits_{i = 1}^N {B_i \exp ( - b_i r^2).} \\
 \end{array}} \right.
\end{equation}

DWF in the coordinate representation $u_{l}(r)$ can be obtained
from the DWF in the momentum representation $u_{l}(p)$ by means of
the Hankel transform \cite{Fujiwara2001}:

\begin{equation}
\label{eq5}
u_l (r) = i^l\sqrt {\frac{2}{\pi }} \int\limits_0^\infty {u_l (p)j_l (pr)dp}
,
\end{equation}

where $j_{0}$\textit{(pr)} and $j_{2}$\textit{(pr)} are Bessel
functions of zero and second order. It was for PEST
\cite{Loiseau1987} that the Hankel transform (\ref{eq5}) can be
used the known momentum parameterisation of DWF.

Certov-Mathelitsch-Moravcsik (CMM) DWF \cite{Certov1987} (the
variant of the potential model in the original paper is designated
as $\psi _{LS}^{6B} )$ have been parameterised by formulas
(\ref{eq1}) with $N$=7.

Obtaining method, approximation interval and knots for DWF are
presented in Table 1.

\begin{center}
\textbf{Table 1}. DWF for nucleon-nucleon interaction potentials
\end{center}

\begin{tabular}{|l|l|l|l|l|l|}
\hline Potential& Obtaining method& Approximation interval (fm)&
Knots for $u(r)$, $w(r)$&
Ref. \\
\hline
Nijm1, Nijm2, Nijm93&
(\ref{eq2})&
0-25&
-/-&
\cite{NPAE2016} \\
\hline Reid93& (\ref{eq2})& 0-25& -/-&
\cite{MPLA2016} \\
\hline Argonne v18& (\ref{eq2})& 0-15& -/-&
\cite{MPLA2016} \\
\hline OBEPC& (\ref{eq1})& -& -/+&
\cite{Machleidt1989} \\
\hline MT& (\ref{eq1})& -& -/-&
\cite{Krutov2007} \\
\hline Paris& (\ref{eq1})& -& -/-&
\cite{Lacombe1981} \\
\hline CDBonn& (\ref{eq1})& -& -/+&
\cite{Machleidt2001} \\
\hline DDM& (\ref{eq3})& -& +/+&
\cite{Platonova2010} \\
\hline fss2& (\ref{eq1})& -& -/+&
\cite{Fujiwara2001} \\
\hline Moscow& (\ref{eq3})& -& +/+&
\cite{Krasnopolsky1985} \\
\hline CMM& (\ref{eq1})& -& -/+&
\cite{Certov1987} \\
\hline PEST& -& -& -/-&
\cite{Loiseau1987} \\
\hline
\end{tabular}

\textbf{3. Asymmetry for Tensor }\textbf{\textit{t}}$_{2j}$\textbf{ and
Vector }\textbf{\textit{t}}$_{1i}$\textbf{ Polarizations}

If the deuteron target is tensor and vector polarized, then the
polarized cross-section is written as follows \cite{Hasell2011}:

\begin{equation}
\label{eq6}
\sigma = \sigma _0 \left[ {1 + \sqrt {\frac{1}{2}} P_{zz} A_d^T + \sqrt
{\frac{3}{2}} P_e P_z A_{ed}^V } \right],
\end{equation}

where target tensor $A_d^T $ and beam-target vector asymmetries
$A_{ed}^V $:

\begin{equation}
\label{eq7}
A_d^T = \frac{3\cos ^2\theta ^\ast - 1}{2}t_{20} (p,\theta _e ) - \sqrt
{\frac{3}{2}} \sin (2\theta ^\ast )\cos \phi ^\ast t_{21} (p,\theta _e ) +
\sqrt {\frac{3}{2}} \sin ^2\theta ^\ast \cos (2\phi ^\ast )t_{22} (p,\theta
_e );
\end{equation}

\begin{equation}
\label{eq8}
A_{ed}^V = \sqrt 3 \left[ {\frac{1}{\sqrt 2 }\cos \theta ^\ast t_{10}
(p,\theta _e ) - \sin \theta ^\ast \cos \phi ^\ast t_{11} (p,\theta _e )}
\right];
\end{equation}

where the angles \textit{$\theta $}$^{\ast }$ and \textit{$\phi
$}$^{\ast }$ define the polarization direction in a frame;
$P_{e}$, $P_{z}$, $P_{zz}$ are the longitudinal polarization of
the electron, the deuteron vector and tensor polarizations.

The polarization of the outgoing (reflected) deuteron can be
measured if the scattering process is analysed in detail.
Differential cross section for the double scattering process is
the following \cite{Gilman2002, Arnold1980}:

\begin{equation}
\label{eq9}
\frac{d\sigma }{d\Omega d\Omega _2 } = \left. {\frac{d\sigma }{d\Omega
d\Omega _2 }} \right|_0 \left[ {1 + \frac{3}{2}ht_{11} A_y \sin \varphi _2 +
\frac{1}{\sqrt 2 }t_{20} A_{zz} - } \right.\left. {\frac{2}{\sqrt 3 }t_{21}
A_{xz} \cos \varphi _2 + \frac{1}{\sqrt 3 }t_{22} (A_{xx} - A_{yy} )\cos
2\varphi _2 } \right],
\end{equation}

where $h=\pm $1/2 - a polarization of the incoming electron beam; \textit{$\phi $}$_{2}$ -
the angle between the two scattering planes; $A_{y}$ and $A_{ij}$ - the vector
and tensor analyzing powers of the second scattering.

The differential cross section for elastic scattering of a
polarized electron beam from a polarized deuteron target is given
by an expression in a laboratory system \cite{Donnelly1986,
Darwish2009}

\begin{equation}
\label{eq10}
\frac{d\sigma }{d\Omega _e }\left( {h;p_z ,p_{zz} } \right) = \Sigma (\theta
,\varphi ) + h\Delta (\theta ,\varphi ),
\end{equation}

where $h$ is a helicity of the incident electron beam; $p_{z}$ and $p_{zz}$
determine the degree of vector and tensor polarizations of the deuteron
target; the angles \textit{$\theta $} and \textit{$\phi $} determine the polarization direction of the
deuteron in the frame.

The first part on the right in a formula (\ref{eq10}) defines a cross-section for an
unpolarized electron but a polarized target deuteron

\begin{equation}
\label{eq11}
\Sigma (\theta ,\varphi ) = \frac{d\sigma _0 }{d\Omega _e }\left[ {1 +
\Gamma (\theta ,\varphi )} \right],
\end{equation}

where $\frac{d\sigma _0 }{d\Omega _e }$ - is an unpolarized
differential cross-section. The value $\Gamma $ contains tensor
deuteron analyzing powers $T_{2j}$:

\begin{equation}
\label{eq12}
\begin{array}{l}
 \Gamma (\theta ,\varphi ) = p_{zz} \left[ {\frac{1}{\sqrt 2 }} \right.P_2^0
(\cos \theta )T_{20} (p^2,\theta _e ) - \frac{1}{\sqrt 3 }P_2^1 (\cos \theta
)\cos \varphi T_{21} (p^2,\theta _e ) + \\
 + \left. {\frac{1}{2\sqrt 3 }P_2^2 (\cos \theta )\cos (2\varphi )T_{22}
(p^2,\theta _e )} \right]. \\
 \end{array}
\end{equation}

The second part on the right in a formula (\ref{eq10}) describes a
helicity-dependent differential cross-section for a polarized
electron beam and a polarized deuteron target and contains vector
deuteron analyzing powers of $T_{10}$ and $T_{11}$
\cite{Darwish2009}:

\begin{equation}
\label{eq13}
h\Delta (\theta ,\varphi ) = h\frac{d\sigma _0 }{d\Omega }p_z \left[
{\frac{\sqrt 3 }{2}P_1 (\cos \theta )T_{10} (p^2,\theta _e ) - \sqrt 3 P_1^1
(\cos \theta )\cos \varphi T_{11} (p^2,\theta _e )} \right].
\end{equation}

The formulas (\ref{eq12}) and (\ref{eq13}) $P_l (x)$and $P_l^m (x)$ determine Legendre
polynomials and associated Legendre polynomials respectively.

The values of the tensor $T_{20}$, $T_{21}$, $T_{22}$ and vector
$T_{10}$, $T_{11}$ of deuteron analyzing powers are determined
formulas (\ref{eq7})-(\ref{eq9}), (\ref{eq12}) and (\ref{eq13})
through form factors such as \cite{Darwish2008, Darwish2009,
Hasell2011} (in equivalent more widely used terms
\cite{Arnold1981, Abbott20001, Gilman2002} tensor $t_{2j}$ and
vector $t_{1i}$ polarizations):

\begin{equation}
\label{eq14}
t_{10} (p,\theta _e ) = - \sqrt {\frac{2}{3}} \frac{\eta }{S}\sqrt {(1 +
\eta )\left( {1 + \eta \sin ^2\left( {\frac{\theta _e }{2}} \right)}
\right)} G_M^2 (p)\mbox{tg}\left( {\frac{\theta _e }{2}} \right)\sec \left(
{\frac{\theta _e }{2}} \right),
\end{equation}

\begin{equation}
\label{eq15}
t_{11} (p,\theta _e ) = \frac{2}{\sqrt 3 S}\sqrt {\eta (1 + \eta )} G_M
(p)\left[ {G_C (p) + \frac{\eta }{3}G_Q (p)} \right]\mbox{tg}\left(
{\frac{\theta _e }{2}} \right),
\end{equation}

\begin{equation}
\label{eq16}
t_{20} (p,\theta _e ) = - \frac{1}{\sqrt 2 S}\left( {\frac{8}{3}\eta G_C
(p)G_Q (p) + \frac{8}{9}\eta ^2G_Q^2 (p) + \frac{1}{3}\eta \left[ {1 + 2(1 +
\eta )\mbox{tg}^2\left( {\frac{\theta _e }{2}} \right)} \right]G_M^2 (p)}
\right),
\end{equation}

\begin{equation}
\label{eq17}
t_{21} (p,\theta _e ) = \frac{2}{\sqrt 3 S\cos \left( {\frac{\theta _e }{2}}
\right)}\eta \sqrt {\eta + \eta ^2\sin ^2\left( {\frac{\theta _e }{2}}
\right)} G_M (p)G_Q (p),
\end{equation}

\begin{equation}
\label{eq18}
t_{22} (p,\theta _e ) = - \frac{1}{2\sqrt 3 S}\eta G_M^2 (p),
\end{equation}

where the factor $S(p,\theta _e ) = A(p) + B(p)\mbox{tg}^2\left(
{\frac{\theta _e }{2}} \right)$ is determined by the structure
functions $A, B$ and the scattering angle \textit{$\theta
$}$_{е}$; the charge $G_{C}(p)$, quadrupole $G_{Q}(p)$ and
magnetic $G_{M}(p)$ form factors contain information about the
electromagnetic properties of the deuteron. The values of the
tensor $t_{20}$ and vector $t_{11}$ polarization are determined by
the form factors $G_{C}(p)$, $G_{Q}(p)$, $G_{M}(p)$ and scattering
angle \textit{$\theta $}$_{е}$, and $t_{21}$ -- by $G_{Q}(p)$,
$G_{M}(p)$ and \textit{$\theta $}$_{е}$. The values of $t_{22}$
and $t_{10}$ depend only on the form factors of the $G_{M}(p)$ and
the scattering angle \textit{$\theta $}$_{е}$.

\textbf{4. Results of calculations}

The values of the angular asymmetry for the components of the
vector $t_{1i}$ and the tensor $t_{2j}$ polarizations for the
first eight potentials in Table 1 (Nijm1, Nijm2, Nijm93, Reid93,
Argonne v18, OBEPC, MT, Paris) are given on Fig. 1-5. These values
at angles 10$^{0}$, 40$^{0}$, 70$^{0}$, 90$^{0}$, 120$^{0}$
calculated by DWF for CD-Bonn, Nijm-93, OSBEP, AV18 and Paris
potentials in \cite{Darwish2009, Darwish2017}. The comparison of
theoretical results and experimental data for vector $t_{1i}$ and
tensor $t_{2j}$ polarization has already been indicated by
\cite{MPLA2016} for Argonne v18 and in \cite{MPLA2016,
Zhaba20171} for Reid93 potentials. Therefore, in this paper, the
the theory and experiment are not compared.

\pdfximage width 100mm {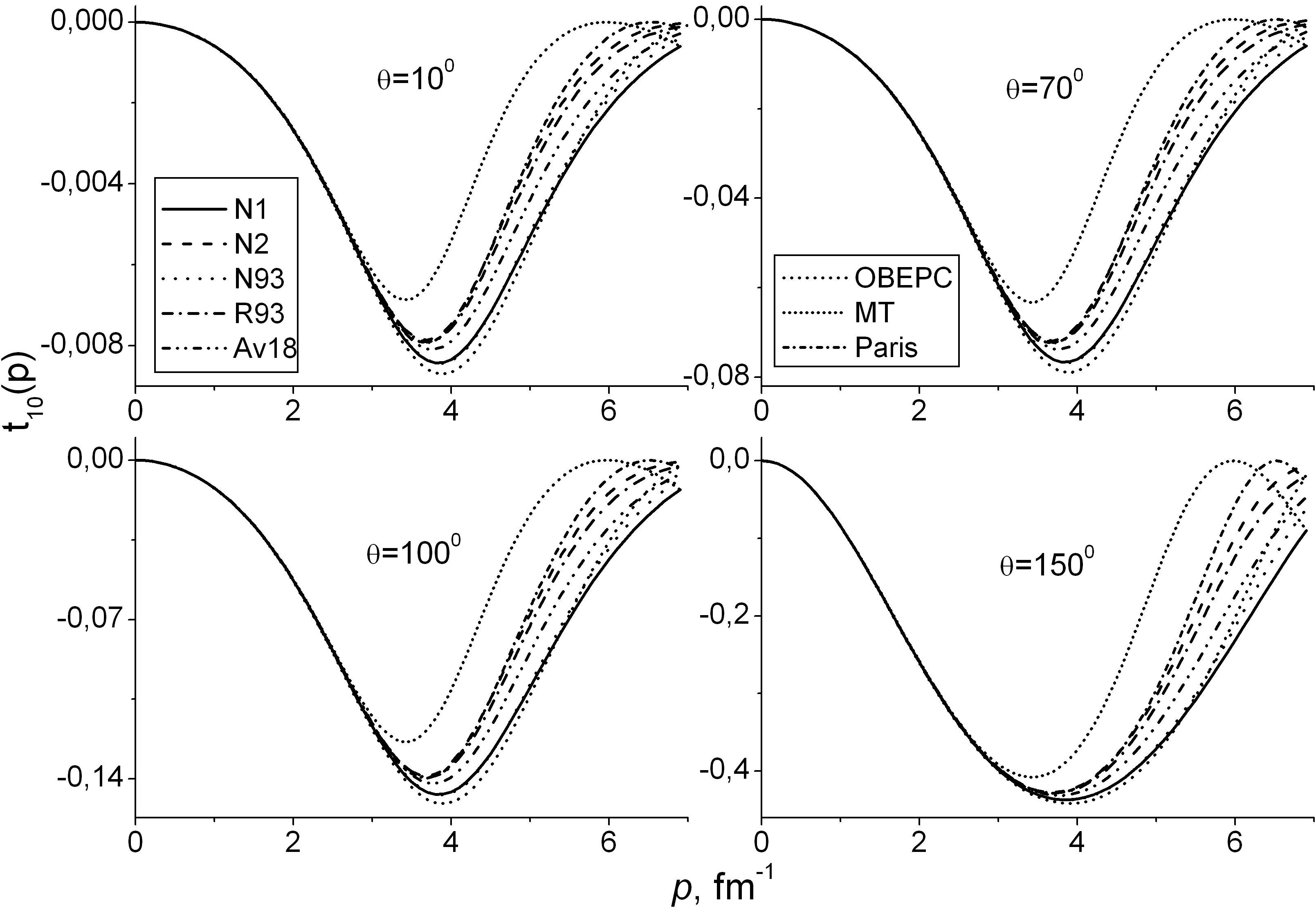}\pdfrefximage\pdflastximage

\textbf{Fig. 1.} Angular asymmetry of vector polarization $t_{10}$

\pdfximage width 100mm {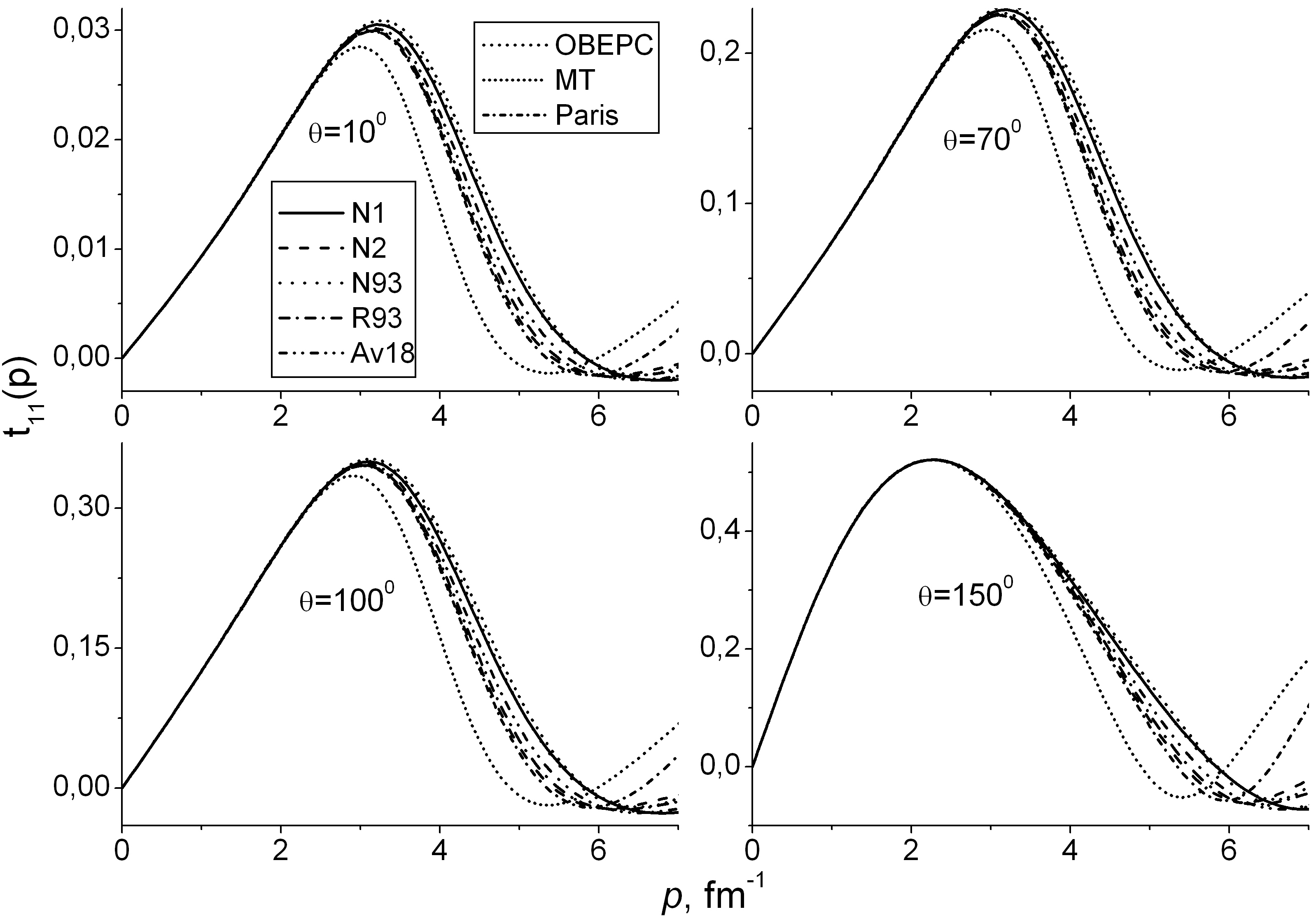}\pdfrefximage\pdflastximage

\textbf{Fig. 2.} Angular asymmetry of vector polarization $t_{11}$

\pdfximage width 100mm {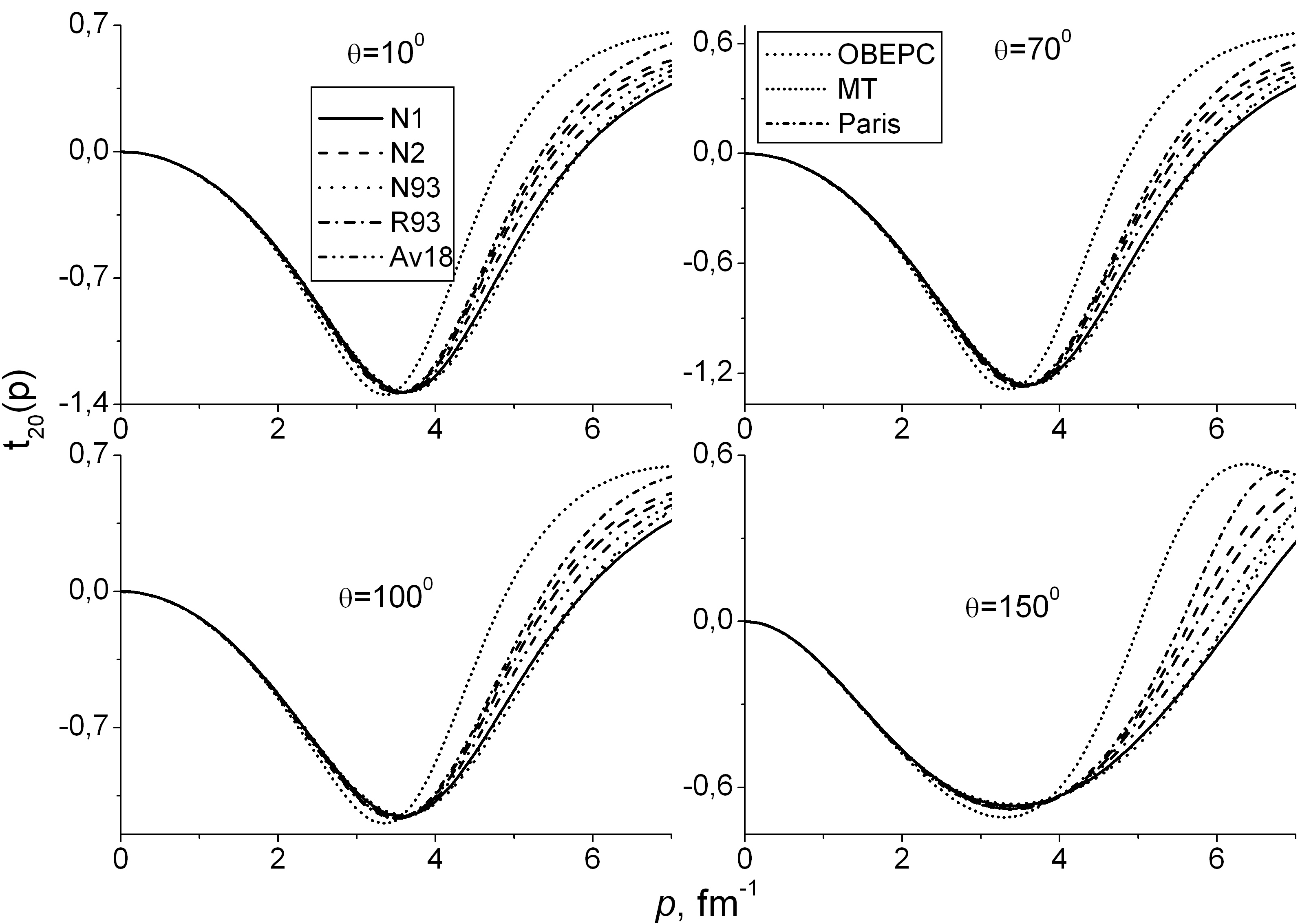}\pdfrefximage\pdflastximage

\textbf{Fig. 3.} Angular asymmetry of tensor polarization $t_{20}$

\pdfximage width 100mm {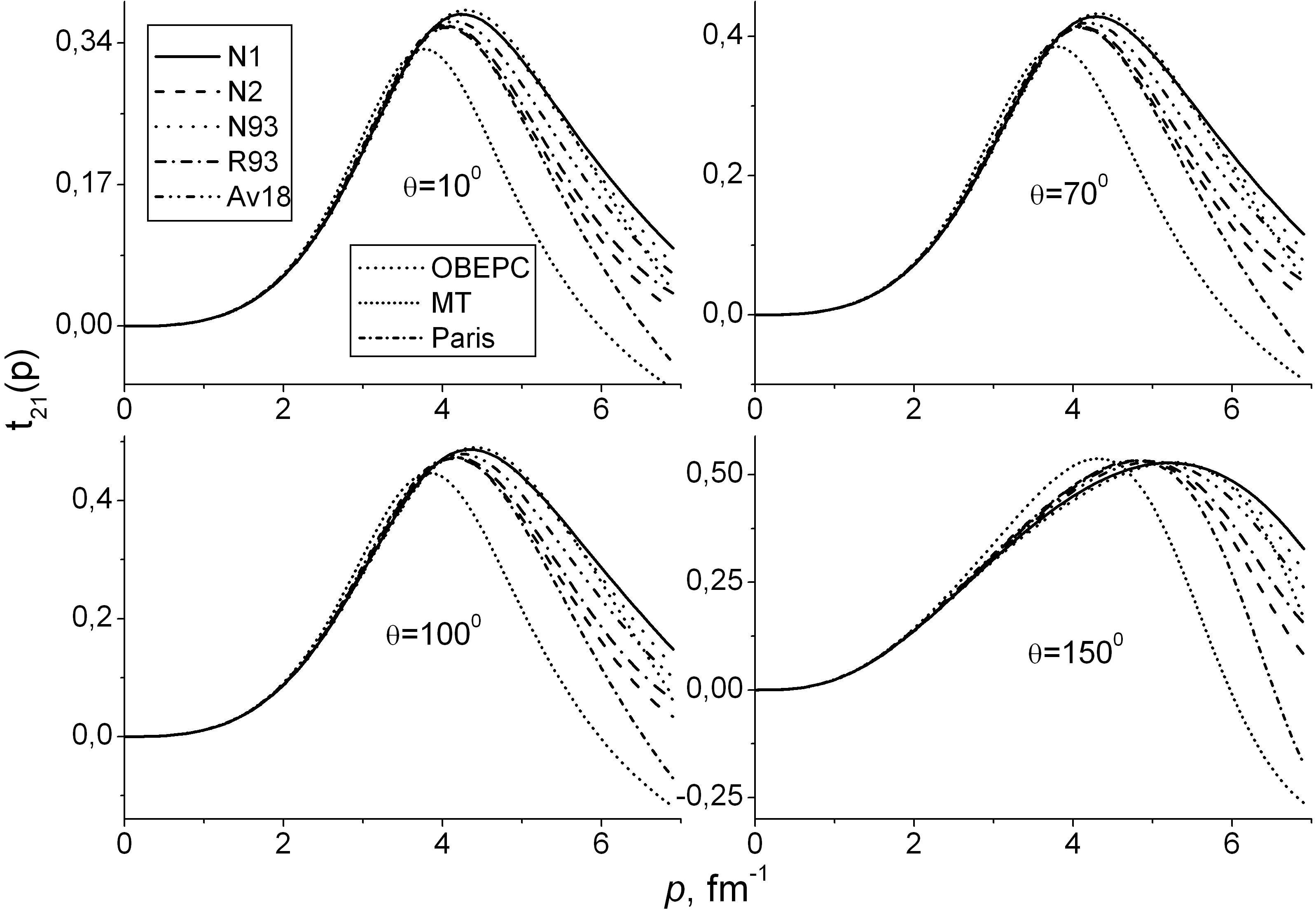}\pdfrefximage\pdflastximage

\textbf{Fig. 4.} Angular asymmetry of tensor polarization $t_{21}$

\pdfximage width 100mm {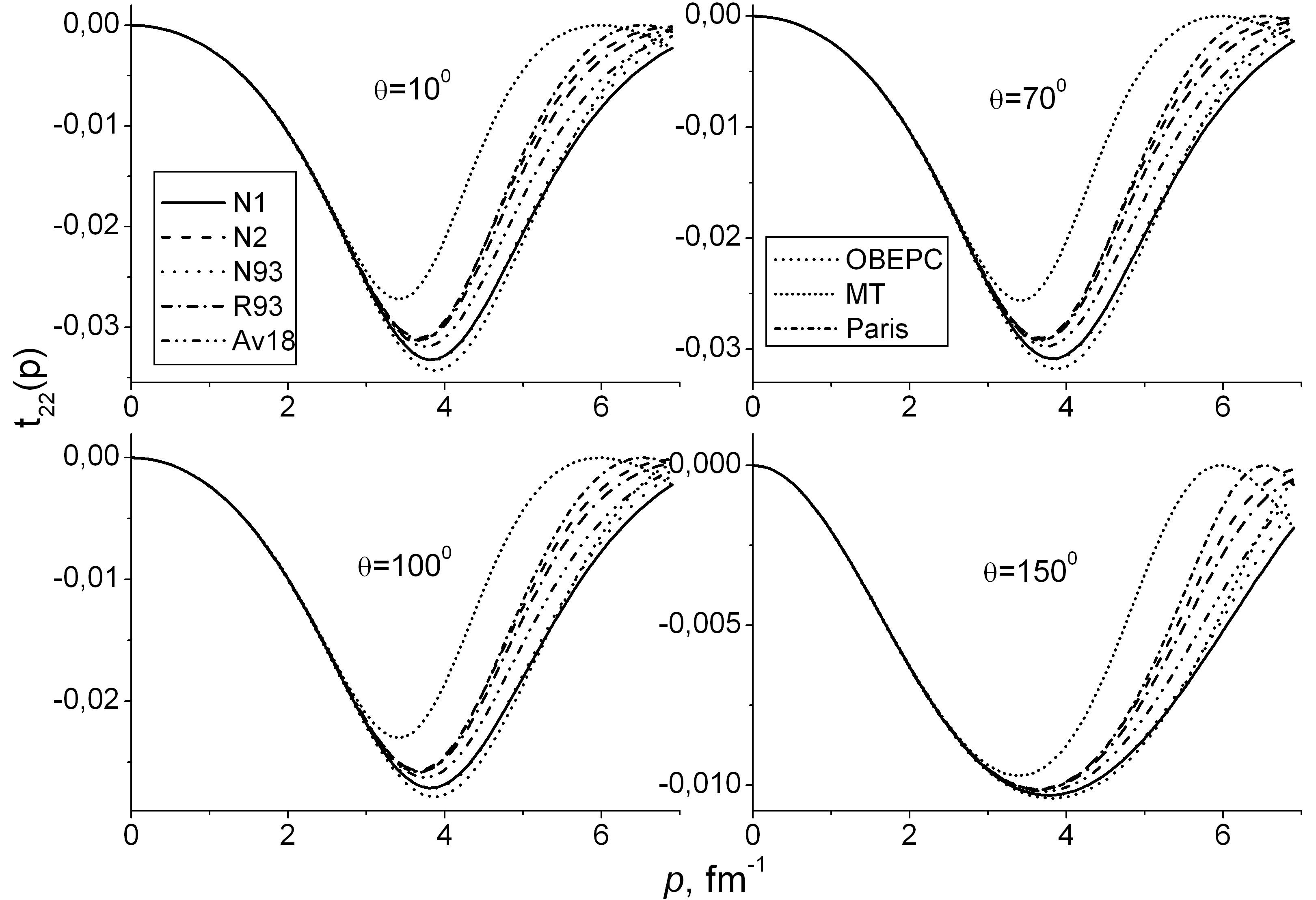}\pdfrefximage\pdflastximage

\textbf{Fig. 5.} Angular asymmetry of tensor polarization $t_{22}$

\pdfximage width 90mm {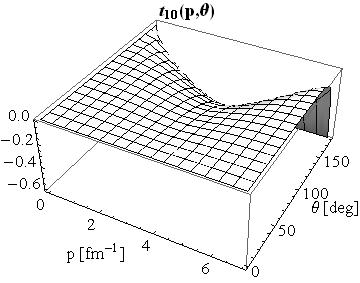}\pdfrefximage\pdflastximage

\textbf{Fig. 6.} Vector polarization $t_{10}$ for Reid93 potential

\pdfximage width 90mm {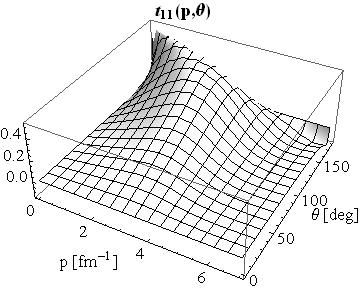}\pdfrefximage\pdflastximage

\textbf{Fig. 7.} Vector polarization $t_{11}$ for Reid93 potential

\pdfximage width 90mm {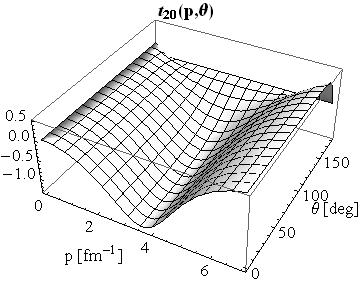}\pdfrefximage\pdflastximage

\textbf{Fig. 8.} Tensor polarization $t_{20}$ for Reid93 potential

\pdfximage width 90mm {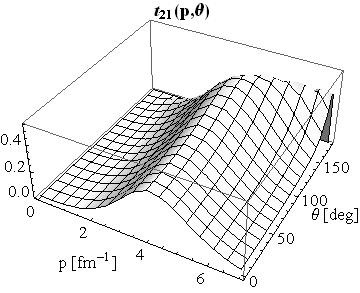}\pdfrefximage\pdflastximage

\textbf{Fig. 9.} Tensor polarization $t_{21}$ for Reid93 potential

\pdfximage width 90mm {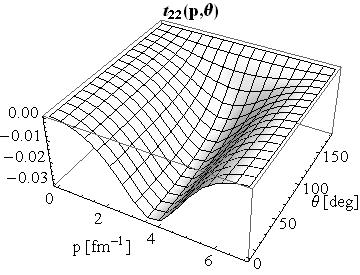}\pdfrefximage\pdflastximage

\textbf{Fig. 10.} Tensor polarization $t_{22}$ for Reid93
potential

The angle and impulse asymmetry of vector and tensor polarization
for DWF (\ref{eq2}) for the Argonne v18 potential in
\cite{IJARPS2018} and for the Reid93 potential in
\cite{Zhaba20171} has been partially studied. These data are
quoted here in Figs. 1-5 for the specified scattering angles. For
other potentials, results are obtained for the first time.

Now let's analyse the calculated components of vector and tensor
polarizations. As it can be seen in Figs. 1 and 2, vector
polarization $t_{1i}$ strongly depend on the scattering angle
\textit{$\theta $}$_{e}$. So, there is an angular asymmetry for
both vector polarization components $t_{10}$ and $t_{11}$. In
addition, at $p$>2.5 fm$^{-1}$, they also depend on the choice of
the potential for nucleon-nucleon interaction. The minimum values
for $t_{10}$ and $t_{11}$ are located in the area $p$=4 і 6 fm$^{
- 1}$ respectively. The comparison of vector polarizations
indicates that the asymmetry $t_{10}$ is less than the asymmetry
of $t_{11}$ at the same angles.

In contrast to the vector polarizations $t_{1i}$, the deuteron
tensor polarization $t_{20}$ (Fig. 3) weakly depends on the
scattering angle. That is, the angular asymmetry for $t_{20}$ will
be slightly intense and weakly sensitive to the scattering angle.
The value of $t_{20}$ will have a constant limit value of 0.3-0.7
in the absence of the angle asymmetry and independence from the
deuteron form factors. The minimum value for $t_{20}$ is in the
pulse region $p$=3-4 fm$^{-1}$.

Due to this behaviour of polarizations $t_{20}$ and $t_{11}$, the
terms "single and double spin asymmetries" are used
\cite{Darwish2017}.

Tensor polarizations $t_{21}$ and $t_{22}$ (Figs. 4 and 5) are
characterized by angular asymmetry, which is more pronounced for
$t_{22}$ at large angles. Asymmetries $t_{21}$ and $t_{22}$ with
increasing scattering angle are described by curves with
progressive convexity and concavity with maximum and minimum at $p
\approx $4 fm$^{-1}$, respectively. The values of $t_{20}$ and
$t_{21}$ give a greater contribution to the absolute value of the
cross-section with the same scattering angles, and $t_{22}$ will
be an order of magnitude smaller.

In next Figs. 6-10 are given calculated values $t_{ij} $ in the $p\theta _e
$ plane for Reid93 potential. This is the angular-momentum dependence of
values vector $t_{10} (p,\theta _e )$ and tensor $t_{2i} (p,\theta _e )$
polarizations in 3D format. The values $t_{10} (p,\theta _e )$ and $t_{11}
(p,\theta _e )$ are typical of flat forms at small angles and asymmetric
momentum dependences at large angles. There is a hump (peak) at 4 fm$^{ -
1}$ in the range of angles 0-180 degrees for tensor polarization $t_{21}$ in
3D format here. There is a pit for the tensor polarizations $t_{20}$ and
$t_{22}$, on the contrary.

\textbf{5.Conclusions}

The novelty of this paper is the complete presentation for angular 
and angular-momentum asymmetries for all components 
of tensor and vector polarizations asymmetries. 
And the resulting angular-momentum asymmetry in the 3D format 
is original for nucleon-nucleon potentials for different models.

In general, it is necessary to detailed and exact study the
influence of the form DWF on the behaviour and changes for
characteristics of processes with the participation of deuteron.
By the way, results of influence of a choice of the analytical
form in coordinate space on calculations of density distribution
in the deuteron and transition density for Reid93 and Moscow
potentials are given in paper \cite{Zhaba20172}.

And further, the obtained results for components of vector and tensor
polarizations can be used to calculate the values of the cross-sections (\ref{eq6}),
(\ref{eq8}), asymmetries (\ref{eq7}), (\ref{eq8}) and other characteristics of processes with the
participation of a deuteron.

Prospective are further studies of the angular-momentum dependence
of values vector $A_B^i (p,\theta )$ and tensor $A_B^{ij}
(p,\theta )$ asymmetries in 3D format \cite{Gakh2012}. Moreover,
it is of interest to conduct the further study of polarization
observables in elastic lepton-deuteron scattering \cite{Gakh2014}
(first of all for the case of spin correlation limit of zero
lepton mass).

It can be added that the obtained results allow us to estimate 
the integral picture for the behavior of tensor and vector polarizations 
in a wider angular, pulse and angular momentum scaling 
for various types of nucleon-nucleon potentials.
In addition, the obtained forms for tensor and vector polarizations 
in 3D format (the results of this paper) can be systematized together 
with the angular-momentum dependence for vector and tensor asymmetries 
in elastic electron-deuteron scattering \cite{Gakh2012}, 
as well as with spin correlation coefficients and tensor asymmetries 
in lepton-deuteron elastic scattering \cite{Gakh2014}. 
To do this, you need to use a set of different types 
of nucleon-nucleon potentials and models.

\end{document}